\documentclass[preprint,12pt]{elsarticle}

\usepackage{amssymb}
\usepackage{color}
\journal{arXiv}

\begin{document}

\begin{frontmatter}



\title{A generalized Holling type II model for the interaction between dextral-sinistral snails and {\it Pareas} snakes}


\author{A. Alonso Izquierdo$^{(a,c)}$, M.A. Gonz\'alez Le\'on$^{(a,c)}$ and M. de la Torre Mayado$^{(b,c)}$}

\address{$^{(a)}$ Departamento de Matematica Aplicada, Universidad de Salamanca, SPAIN \\
$^{(b)}$ Departamento de Fisica Fundamental, Universidad de Salamanca, SPAIN \\$^{(c)}$ IUFFyM, Universidad de Salamanca, SPAIN}

\begin{abstract}

\textsl{Pareatic} snakes possess outstanding asymmetry in the mandibular tooth number, which has probably been caused by its evolution to improve the feeding on the predominant dextral snails. Gene mutation can generate chiral inversion on the snail body. A sinistral snail population can thrive in this ecological context. The interactions between dextral/sinistral snails and \textsl{Pareas} snakes are modeled in this paper by using a new generalized functional response of Holling type II. Distinct \textit{Pareas} species show different bilateral asymmetry degrees. This parameter plays an essential role in our model and determines the evolution of the populations. Stability of the solutions is also analyzed for different regimes in the space of parameters.
\end{abstract}


\end{frontmatter}


\section{Introduction}
\label{Introduction}

Bilateral symmetry in external appearance is a common feature of free-living animals that is sometimes broken because of the functional advantages that can be derived from the presence of asymmetry in some concrete external organs. A fascinating example of this fact is given by pareatic snakes. \textit{Pareas} is a genus of Asian snail-eating specialist snakes, see \cite{Hoso1} and references therein. Snail species have predominantly dextral (clockwise coiled) shells \cite{Gittenberger, Schil, Vermeij}, so most pareatid snakes have evolved asymmetry in mandibular tooth number \cite{Hoso1}. This physiological adaptation facilitates snail body extraction from the shell \cite{Gotz1} and leads to the specialization of these predators in feeding on dextral snails. The paradigm of mandibular asymmetry corresponds to the \textit{Pareas iwasakii} snakes, which have approximately 17 teeth on the left side and 25 teeth on the right side \cite{Hoso1,Hoso2}. This is the extreme case but almost all the 14 different pareatine species \cite{Hoso1, You} involve distinct degrees of mandibular tooth asymmetry, that in principle reflects dietary specialization on dextral snails.

On the other hand, certain snail gene mutations can give rise to a sinistral (counterclockwise coiled) snail population. Copulation between dextral and sinistral snails is usually strongly impeded by genital and behavioural mismatches \cite{Hoso2,Asami1998,Ueshima2003}. These circumstances lead to instant snail speciation, such that dextral/sinistral snails can be interpreted as different populations from a mathematical point of view. Usually the sinistral snail population remains small due to competence with the dextral variant. However, in presence of dextral snail-eating specialist snakes the survival advantage of left-handed snails allows this population to thrive. This mechanism opens the possibility that sinistral snail population replaces the dextral snail population. Indeed, it is a fact that left-right reversal has evolved many times, especially in terrestrial snails \cite{Vermeij,Robertson1993}.

In the extreme case of the \textit{Pareas iwasakii} snakes, \textit{Satsuma} snails constitute its fundamental feeding. The interactions between this snake species and the dextral/sinistral \textit{Satsuma} variants have been thoroughly studied by Hoso and his collaborators in references \cite{Hoso2, Hoso3}. In the lab experiments carried out by these researchers none of the dextral snails survived snake predation where as 87.5 \% of sinistral snails survived predation. Due to the right handedness of the striking direction in the hunting process, the snake can rarely grasp a sinistral snail. The ratio between the hunting success rates on sinistral/dextral snails is approximately equal to 0.12. On the other hand, it is also interesting to remark that the mandibular asymmetry is not always accompanied by chirality specialization, as shown in diverse studies on the \textit{Pareas Carinatus} species \cite{Danai}. \textit{Pareas Carinatus} snakes exhibit a relatively week dental asymmetry in the genus but, in addition, they recognize prey handedness. This ability allows this type of snakes to strike by tilting its head either leftward or rightward depending on the snail chirality, as has recently been shown in \cite{Danaisawadi2016}. As a consequence, these snakes prey both dextral and sinistral snails with similar efficiency.

Returning to the particular interactions between pareatic snakes and \textit{Satsuma} snails, it is important to bring attention to the geographical distribution of the habitats that are shared by these species. It has been well documented the existence of South-East Asian islands where (1) only the dextral \textit{Satsuma} snail variant inhabits, (2) only the dextral and sinistral \textit{Satsuma} snail populations coexist, (3) the snake and sinistral/dextral \textit{Satsuma} snail populations are present and (4) only sinistral \textit{Satsuma} snails arise. For example, \textit{Satsuma} snails cohabit with four different \textit{Pareas} snake species in Taiwan Island, see Figure 5 in reference \cite{Hoso2}. Obviously, the coexistence of pareatic snakes and dextral snails is the standard for other snail species and Asian regions. Taking into account all the previous scheme, it seems to us that the construction of a mathematical model which can be used to unveil the fate of the three involved populations is worthwhile.

The goal of this paper is threefold. Firstly, a novel mathematical model is constructed to describe the interactions between one-predator and two-prey variant populations. In order to accomplish this task we shall use the same assumptions on the hunting habits of predators that are employed in Holling's type II models \cite{Holling1, Holling2, Holling3, May, Maynard}. In this sense our model can be understood as a generalization of this type of models to the one-predator two-prey population context. It is clear that the dextral/sinistral snail populations compete by the same resources and are described by the same ecological parameters. The biased relationship between the pareatic snakes and the two snail variants will be responsible of the asymmetry in our model. Therefore, the ratio between the depredatory efficiencies on the dextral/sinistral snails (which depends on the \textit{Pareas} snake species) plays an essential role in our model. Secondly, the mathematical translation of the previous hypotheses leads to a system of three ordinary differential equations. The stability of the stationary points of this system is discussed. A qualitative analysis of the solutions depending of the model parameters is also considered. Thirdly, the previously mentioned analytical approach is used to classify the possible ecological scenarios. It will be shown that chirality reversal in the snail population induced by the snake specialization is dictated by the population dynamics in many parameter ranges but if the snake specialization is too strong there is also room for snake extinction with coexistence of dextral/sinistral snail populations.

The structure of the paper is as follows: In Section 2 the mathematical model is constructed. Section 3 is devoted to analyze several particular and limiting cases where the model reduces to simpler well-known systems. The description of the model dynamics is presented in Section 4. The local stability of the different stationary points and a qualitative analysis of the solutions is also considered in this section. The possible final scenarios depending on the ecological parameters are classified. A brief discussion and conclusions are presented in Section 5. Finally, an Appendix with technical details and proofs of mathematical results obtained in Section 4 has been added.

\section{Construction of the model}

\label{Themodel}

Prey population densities (population per unit area) of dextral and sinistral snails will be denoted by the functions $X_1(t)$ and $X_2(t)$ respectively, whereas $Y(t)$ will be used to represent the predator population density of the \textit{Pareas} snakes in a closed homogeneous environment. The octant
$\mathbb{E}=\{(X_1,X_2,Y)\in \mathbb{R}^3 : X_1\geq 0 ,X_2\geq 0, Y \geq 0\}$ of the phase space $\mathbb{R}^3$ defines the region of ecological interest. A two-prey one-predator population model with different functional responses $\Phi_i(X_1,X_2)$ for both types of prey follows the form:
\begin{eqnarray}
\frac{dX_1}{dt} &=& f_1(X_1,X_2) - \Phi_1(X_1,X_2) Y \hspace{3cm}, \nonumber \\
\frac{dX_2}{dt} &=& f_2(X_1,X_2) - \Phi_2(X_1,X_2) Y  \hspace{3cm},  \label{ecu1} \\
\frac{dY}{dt} &=& -s  Y  +  \left( b_1\, \Phi_1(X_1,X_2)+ b_2 \, \Phi_2(X_1,X_2)\right)  Y \hspace{0.4cm}, \nonumber
\end{eqnarray}
where the functions $f_i(X_1,X_2)$, $i=1,2$, describes the evolution of the prey populations without predators. In order to set the undetermined functions in the system of differential equations (\ref{ecu1}) the following assumptions are considered:

\vspace{0.2cm}

\noindent \textbf{A.} In the absence of predators, a logistic growth for the whole snail population is assumed. As previously mentioned, the dextral/sinistral snails considered in this paper are variants of the same species, only differing in body chirality. Therefore, the same intrinsic growth rate $r$ is conjectured for the two snail populations $X_1(t)$ and $X_2(t)$. The linear dependence on these variables in the equations (\ref{ecu1}) must be uncoupled because of reproductivity incompatibility (as previously noticed) \cite{Hoso1}. Moreover, the same ecosystem is shared by these organisms, so a common carrying capacity $K$ (per unit area) relative to the sum of both populations must be introduced. All the individuals compete for the same food. The previous observations lead to the expressions
\begin{eqnarray}
f_1(X_1,X_2) &=& r X_1 \, \left( 1-\frac{X_1+X_2}{K}\right) - g_1(X_1,X_2) \hspace{0.3cm}, \label{f1} \\
f_2(X_1,X_2) &=& r X_2 \, \left( 1-\frac{X_1+X_2}{K}\right) - g_2(X_1,X_2) \hspace{0.3cm}, \label{f2}
\end{eqnarray}
where $g_i(X_1,X_2)$ measure a residual interspecific competition between the two kind of snails. From our previous analysis, it is assumed that these functions vanish, or at least, they can be neglected in a first approach, i.e., $g_i(X_1,X_2) \approx 0$ with $i=1,2$.

\hspace{0.2cm}

\noindent \textbf{B.} The functional responses $\Phi_1(X_1,X_2)$ and $\Phi_2(X_1,X_2)$ measure the number of preys consumed by a predator per unit time. The subscript $i=1,2$ distinguishes, respectively, if the prey is a dextral or a sinistral snail. The intake of any type of snail is equally beneficial for the pareatic snake population growth. This fact implies that the parameters $b_i$ (the efficiency with which predators convert consumed preys into new snake offsprings) in (\ref{ecu1}) are equal, that is, $b_1=b_2=b$. Now, we will construct the terms $\Phi_i(X_1,X_2)$ by following the same ecological arguments employed in the deduction of the Holling's type II functional response for one-prey one-predator population models \cite{Holling1, Holling2, Holling3}.

Let $T$ denote the time devoted by a predator to the process of hunting and handling preys, which it is assumed to be constant in the snake daily life. This amount of time $T$ is the sum of two different time intervals,
\begin{equation}
T=T_S+T_H \hspace{0.4cm}, \label{totaltime}
\end{equation}
where $T_S$ is the time spent in searching, pursuing and hunting preys, whereas $T_H$ represents the time employed in \lq\lq handling" preys. $T_H$ can be understood as the time lag between the successful hunting of a prey and the disposition of the predator to start again the process of capture. If $A$ denotes the area supervised by the predator per unit time, $A\, T_S\, X_i$ will be the number of preys of type $i$ detected by one predator in the time interval $T_S$. If $e_i$, $i=1,2$ represent the efficiencies of the predator when capturing a prey of type $i$ (quotient between the number of successful and total attacks), then
\begin{equation}
\delta X_i= A\, T_S\, X_i\, e_i\, , \qquad i=1,2 \label{deltaX}
\end{equation}
totals the number of preys of type $i$ captured by a predator in the searching time $T_S$. From (\ref{deltaX}) the following relation
\[
T_S=\frac{\delta{X}_1}{e_1 A X_1} =\frac{\delta{X}_2}{e_2 A X_2}
\]
can be directly obtained, which means that
\begin{equation}
\frac{\delta{X}_1}{e_1 X_1} = \frac{\delta{X}_2}{e_2 X_2} \hspace{0.3cm}.
\label{tiempoS}
\end{equation}
On the other hand, the handling time $T_H$ must be proportional to the number of hunted preys $\delta{X}_1+\delta{X}_2$, i.e.
\begin{equation}
T_H= t_h (\delta{X}_1+\delta{X}_2)\label{th} \hspace{0.3cm} ,
\end{equation}
where the proportionality constant $t_h$ is the handling time per captured prey unit. This per capita handling time $t_h$ is obviously independent of the snail chirality. Plugging (\ref{tiempoS}) and (\ref{th}) into (\ref{totaltime}), the number of preys of type $i$ consumed by a predator in the total time $T$ is written as
\begin{equation}
\delta{X}_i =  \frac{e_i A T \,X_i}{1+t_h A (e_1 X_1+e_2 X_2)}\  ,\quad i=1,2\label{deltas}
\end{equation}
in terms of the ecological features measured by the parameters $A$, $e_i$, $t_h$ and $T$. Finally, the functional responses $\Phi_1(X_1,X_2)$ and $\Phi_2(X_1,X_2)$ are proportional to $\delta{X}_1$ and $\delta{X}_2$ respectively,
\begin{equation}
\Phi_1(X_1,X_2)\, =\, a\, \delta{X}_1\  ,\quad \Phi_2(X_1,X_2)\, =\, a\, \delta{X}_2 \hspace{0.3cm}, \label{func}
\end{equation}
where $a$ is the predator per capita prey consumption rate. Obviously the nutritional needs of the snake are independent of the prey type, so the parameter $a$ has been considered equal for the dextral/sinistral snail preys. Substituting (\ref{f1}), (\ref{f2}), (\ref{deltas}) and (\ref{func}) into (\ref{ecu1}) leads to the system of differential equations
\begin{eqnarray}
\frac{dX_1}{dt} & =&\displaystyle r \, X_1 \Big( 1-\frac{X_1 +X_2 }{K} \Big) - \frac{a\, e_1 \,A T \,X_1 \,Y}{1+t_h A (e_1X_1 +e_2X_2)} \hspace{0.3cm}, \nonumber \\
\frac{dX_2}{dt} & =&\displaystyle r \, X_2  \Big( 1-\frac{X_1 +X_2 }{K} \Big) - \frac{a\, e_2\, A T \,X_2 \,Y}{1+t_h A (e_1X_1 +e_2X_2 )} \hspace{0.3cm}, \label{system}\\
\frac{dY}{dt}&=& \displaystyle-s Y  +\, b \, a\, A\, T\, \frac{ \left( e_1  \,X_1+\, e_2\, X_2\right) \, Y }{1+t_h A (e_1 X_1 +e_2X_2)} \hspace{1.8cm}, \nonumber
\end{eqnarray}
which determines our dextral/sinistral snail prey \textit{Pareas} snake predator population model. In Table 1 the parameters included in (\ref{system}) are summarized.

\begin{table}[h]
\centerline{\small
\begin{tabular}{cl} \hline
\textsc{Parameter} & \textsc{Description} \\ \hline
$r$ & \textit{Intrinsic snail growth rate.}  \\
$s$ & \textit{Intrinsic Pareas snake mortality rate in absent of snails.}  \\
$K$ & \textit{Snail carrying capacity (per unit area).} \\
$a$ & \textit{Pareas snake per capita snail consumption rate.} \\
$T$ & \textit{Total hunting and handling time employed by a Pareas snake.} \\
$t_h$ & \textit{Handling time per captured snail employed by a snake.} \\
$A$ & \textit{Area supervised by a Pareas snake per unit time.} \\
$b$ & \textit{Pareas snake consumption efficiency.} \\
$e_1$ & \textit{Pareas snake hunting success rate on dextral snails.} \\
$e_2$ & \textit{Pareas snake hunting success rate on sinistral snails.} \\ \hline
\end{tabular}
}
\caption{Description of the ecological parameters introduced in the system (\ref{system}).}
\end{table}

In order to study the system of differential equations (\ref{system}) it is convenient to introduce non-dimensional variables
\begin{equation}
\tau=rt\ ,\quad x_i=\frac{X_i}{K}\  ,\quad y=\frac{a e_1 A T}{r} Y \hspace{0.3cm},  \label{adim}
\end{equation}
together with non-dimensional coefficients
\[
\epsilon_i=t_h  A K\, e_i\  , \quad \sigma=\frac{s}{r}\  ,\quad \beta=\frac{bT}{t_h}\  ,\quad \alpha=\frac{e_2}{e_1} \hspace{0.3cm},
\]
in such a way that the system (\ref{system}) reduces to
\begin{eqnarray}
\frac{dx_1}{d\tau} & =&\displaystyle  x_1 ( 1-x_1-x_2 ) - \frac{x_1 \,y}{1+ \epsilon_1 (x_1 + \alpha  x_2)} \hspace{0.3cm}, \nonumber \\
\displaystyle\frac{dx_2}{d\tau} & =& x_2  ( 1-x_1-x_2) - \frac{ \alpha \,x_2 \,y}{1+ \epsilon_1 (x_1 + \alpha x_2)}\hspace{0.3cm}, \label{system2} \\
\frac{dy}{d\tau} &=& \displaystyle \left( \beta-\sigma\right)\, y\, -\, \frac{\beta\, y}{1+\epsilon_1 (x_1 +\alpha x_2)} \hspace{1.3cm}, \nonumber
\end{eqnarray}
that depends only on four non-negative non-dimensional parameters: $\alpha$, $\beta$, $\sigma$ and $\epsilon_1$. Now, $\mathbb{E}=\{(x_1,x_2,y)\in \mathbb{R}^3 : x_1\geq 0 ,x_2\geq 0, y\geq 0\}$. A first constraint in the non-dimensional parameters can be obtained by ecological arguments. From the third equation in the system (\ref{system2}) a viable predator population must comply with the inequality
\begin{equation}
\beta\, >\, \sigma \hspace{0.2cm}. \label{range1}
\end{equation}
Compliance with (\ref{range1}) requires that the physiological conditions of the \textit{Pareas} snakes be favorable enough to survive in the ecosystem. Hunting time $T$ and consumption efficiency $b$ must be large enough and the intrinsic mortality rate $s$ and handling time $t_h$ must be small enough so that the quotient $(st_h)/(bT)$ be less that the intrinsic snail growth rate. Otherwise the snake population will become extinct.

The physiological bilateral asymmetry of the pareatic snakes, manifested in the distinct hunting efficiencies on dextral/sinistral snails, is hold in the non-dimensional parameter $\alpha$. By convention, it is assumed that the predator is better adapted to hunting the prey 1 than prey 2. In our context, {\it Pareas} snakes are more efficient in hunting dextral snails than sinistral ones, i.e., $e_1 \geq e_2$. Therefore, the non-dimensional parameter $\alpha$ is confined to the range
\begin{equation}
0 \leq \alpha \leq 1 \hspace{0.4cm}. \label{alpharange}
\end{equation}
The conditions (\ref{alpharange}) and (\ref{range1}) will be assumed from now on. The value $\alpha=1$ corresponds to \textit{Pareas} snake species which exhibit bilateral symmetry and are equally competent to hunt dextral or sinistral snails whereas $\alpha=0$ indicates total ineptitude to grasp sinistral snails.

Lab experiments with \textit{Pareas iwasakii} snakes and dextral/sinistral \textit{Satsuma} snails carried out by Hoso and collaborators allow us to obtain the value of $\alpha$ for this snake species \cite{Hoso1}. Taking into account that none of the dextral snails survived snake predation where as 87.5 \% of sinistral snails survived predation the value of $\alpha$ is approximately equal to $\alpha=0.12$. The asymmetry index (defined as $(R-L)\times 100/(R+L)$ where $R$ and $L$ are the tooth numbers on the right and left mandibles, respectively) is approximately 17.5 in this case. \textit{Pareas atayal} shows an asymmetry index at least so high than \textit{Pareas iwasakii}, so $\alpha$ is estimated to value in the range $\alpha\in [0.1,0.2]$. This value interval for the parameter $\alpha$ is also considered valid for \textit{Pareas macularius}. A slightly less asymmetry index is exhibited by \textit{Pareas formosensis} and \textit{Pareas chinensis}, such that the value of $\alpha$ is assessed as $\alpha\in [0.2,0.4]$. On the other hand, \textit{Pareas carinatus} although has an asymmetry index of $10.8$ has learnt to recognize prey handedness, which allows this type of snakes to adjust its attack to the snail chirality \cite{Danaisawadi2016}. Lab experiments for this type of snakes indicate that its hunting efficiencies on dextral and sinistral snails have similar values, thus, $\alpha\approx 1$ in this case. \textit{Pareas boulengeri}, \textit{Pareas hantoni}, \textit{Pareas margaritophorus}, \textit{Pareas nuchalis}, \textit{Pareas stanleyi} and \textit{Pareas komaii} present asymmetry index in the interval $[7.5,12.5]$, similar to the \textit{Pareas carinatus}, although it is uncertain if this snake species are endowed with chirality recognition ability. A cautious estimation of the parameter $\alpha$ for these cases is given by the value range $\alpha\in [0.65,0.95]$.

Although the rest of parameters are difficult to assess and surely depending on the particular environment where the populations are settled, the analysis of the solutions of (\ref{system2}) can provide us with relevant information about the evolution of the snail and snake population where a sinistral variant is introduced in the ecosystem. Clearly, the final scenarios of this evolution depends on the parameter $\alpha$, as will be shown later.

\section{Particular and Limiting cases}
\label{Cases}

\noindent We begin the analysis of the system (\ref{system2}) by displaying a plethora of particular or limiting cases, for which the general model reduces to diverse well known predator-prey models:

\vspace{0.2cm}

\noindent \textbf{1.} If there is no second prey population, $x_2(\tau)\equiv 0$, the equations (\ref{system2}) becomes
\begin{equation}
\frac{dx_1}{d\tau} =  x_1 ( 1-x_1 ) - \frac{x_1 \,y}{1+ \epsilon_1 x_1 } \hspace{0.4cm},\hspace{0.4cm}
\frac{dy}{d\tau} = \left( \beta-\sigma\right)\, y\, -\, \frac{\beta\, y}{1+\epsilon_1 x_1 } \hspace{0.3cm},
\label{particular1}
\end{equation}
which corresponds to a standard Holling's type II one-prey one-predator model. This situation is replicated when the first prey population vanishes, $x_1(\tau)\equiv 0$, $\forall \tau \in {\mathbb R}$. This condition leads to the differential equations
\begin{equation}
\frac{dx_2}{d\tau}  = x_2  ( 1-x_2) - \frac{ \alpha \,x_2 \,y}{1+ \epsilon_1  \alpha x_2}\hspace{0.4cm},\hspace{0.4cm}
\frac{dy}{d\tau} = \left( \beta-\sigma\right)\, y\, -\, \frac{\beta\, y}{1+\epsilon_1 \alpha x_2} \hspace{0.3cm}.
\label{particular1b}
\end{equation}
The ecological coefficients which enter in the systems (\ref{particular1}) and (\ref{particular1b}) are, however, different due to the fact that the interactions between the predator and the two types of preys quantitatively differ.

\hspace{0.2cm}

\noindent  \textbf{2.} If the ecosystem lacks predators, $y(\tau) \equiv  0$, (\ref{system2}) reduces to
\begin{equation}
\frac{dx_1}{d\tau}  = x_1 ( 1-x_1-x_2 ) \hspace{0.4cm},\hspace{0.4cm}
\frac{dx_2}{d\tau}  = x_2  ( 1-x_1-x_2)\hspace{0.4cm}, \label{particular2}
\end{equation}
which can be understood as a population model for two species with identical characteristics in mutual competition.

\hspace{0.2cm}

\noindent \textbf{3.} An interesting case emerges when the handling time $t_h$ is considered to be negligible with respect to the rest of characteristic times of the problem. If the limit $t_h\to 0$ is introduced in the system (\ref{system2}) the following relations
\begin{eqnarray}
\frac{dx_1}{d\tau} & =& x_1 ( 1-x_1-x_2 ) - x_1 \,y \hspace{0.5cm}, \nonumber \\
\frac{dx_2}{d\tau} & =&  x_2  ( 1-x_1-x_2) - \alpha \,x_2 \,y \hspace{0.2cm}, \label{lotka} \\
\frac{dy}{d\tau} &=& \left( -\sigma +\bar{\beta} (x_1+\alpha x_2)\right) \, y  \hspace{0.5cm}, \nonumber
\end{eqnarray}
are found. In other words, a generalized Lotka-Volterra model for one predator and two preys is obtained from our model (\ref{system2}) when predators has the ability of instantaneously digesting its preys. Indeed, this observation reveals a handicap of the Lotka-Volterra type models. A redefinition of the non-dimensional parameters
\[
\bar{\beta}\, =\, \epsilon_1 \beta = be_1 ATK
\]
has been used in (\ref{lotka}) in order to avoid the singularity that appears in the definition of $\beta$ for this limiting case. This type of models has been studied in \cite{Goh1977,Hutson1983} and references therein.

\hspace{0.2cm}

\noindent \textbf{4.} If pareatic snakes exhibit bilateral symmetry, i.e., they lack chirality specialization, then $\alpha=1$ and (\ref{system2}) can be written as
\begin{equation}
\frac{dx}{d\tau} = x( 1-x ) - \frac{x\,y}{1+ \epsilon  x} \hspace{0.3cm},\hspace{0.3cm} \frac{dy}{d\tau} =\left( \beta-\sigma\right)\, y\, -\, \frac{\beta\, y}{1+\epsilon x} \hspace{0.4cm}, \label{particular4}
\end{equation}
where $x=x_1+x_2$ is the sum of the dextral and sinistral snails, which behave as an unique population in this particular case. Equations (\ref{particular4}) determine a Holling type II one-prey one-predator population model, similar to case 1 although with different parameter values. \textit{Pareas Carinatus} can be described by this class of models.

\hspace{0.2cm}

\noindent \textbf{5.} In the opposite case, where the snakes are only capable of hunting dextral snails, the hunting success rate $e_2$ vanishes and thus $\alpha=0$. In this hypothetical situation the system (\ref{system2}) becomes
\begin{eqnarray}
\frac{dx_1}{d\tau}  &=& x_1 ( 1-x_1-x_2 ) - \frac{x_1 \,y}{1+ \epsilon_1 x_1 } \hspace{0.5cm},  \nonumber \\
\frac{dx_2}{d\tau}  &=& x_2  ( 1-x_1-x_2) \hspace{2.6cm},  \label{particular5} \\
\frac{dy}{d\tau} &=& \left( \beta-\sigma\right)\, y\, -\, \frac{\beta\, y}{1+\epsilon_1 x_1 } \hspace{1.5cm},  \nonumber
\end{eqnarray}
that represents a one-predator one-prey population model with the presence of a competitor for the prey, see \cite{Mukherjee} and references therein.

\section{Description of the model dynamics}

In this section we shall analyze the stationary points of the equations (\ref{system2}) and their local stability properties. Afterward, global features of the population dynamics determined by (\ref{system2}) are investigated.

It is not difficult to check that all the stationary points of the system (\ref{system2}) are necessarily located on Coordinate planes, that is, there are no steady states of (\ref{system2}) where the three populations coexist. Therefore, the stationary points of the model always involve the lack of at least one of the species. The absence of population determines the trivial stationary point $P_0\equiv (0,0,0)$, which lacks ecological interest. On the $x_1-y$ coordinate plane,
\begin{equation}
P_1\equiv \Big( \frac{\sigma}{\epsilon_1(\beta-\sigma)} , 0 , \frac{\beta}{\beta-\sigma} \left( 1-\frac{\sigma}{\epsilon_1 (\beta-\sigma)}\right) \Big) \label{p1}
\end{equation}
corresponds to an analytical steady solution of (\ref{system2}) where the sinistral snail population is absent. The main goal in this paper is to study the effect of introducing a mutant population (sinistral snails) in a mature ecosystem inhabited by a prey population (dextral snails) and a predator population (\textit{Pareas} snakes). This ecological assumption implies that the non-null components in (\ref{p1}) must be positive,  which leads to the following inequality
\begin{equation}
\epsilon_1 > \frac{\sigma}{\beta-\sigma} \hspace{0.2cm}. \label{inequation01}
\end{equation}
The restriction (\ref{inequation01}) on the model parameters guarantees the possible presence of a coexisting dextral snail and \textit{Pareas} snake populations at the initial time for the general model. From now on the compliance of (\ref{inequation01}) will be assumed for our mathematical model.

On the $x_1-x_2$ coordinate plane, the one-parametric family of steady states
\begin{equation}
R_{12}^{(\mu)} \equiv (\mu,1-\mu,0)\hspace{0.6cm} \mbox{with} \hspace{0.6cm} \mu\in [0,1] \label{r12}
\end{equation}
emerges, see Figure 1. The stationary points $R_{12}^{(\mu)}$ given by (\ref{r12}) correspond to the situation where the total snail population reaches the carrying capacity whereas the predator population is absent. The value of the parameter $\mu$ in (\ref{r12}) determines the percentage of dextral/sinistral snail populations in the constant solution $R_{12}^{(\mu)}$.

Finally, the steady state arising on the $x_2-y$ plane
\begin{equation}
P_2\equiv \Big( 0, \frac{\sigma}{\alpha \epsilon_1(\beta-\sigma)}, \frac{\beta[\beta \alpha \epsilon_1-(1+\alpha\epsilon_1)\sigma]}{\alpha^2 \epsilon_1 (\beta-\sigma)^2} \Big)
\label{p2}
\end{equation}
completes the stationary point catalogue of the system (\ref{system2}). Assuming the condition (\ref{inequation01}) two different ecological situations are distinguished by the sign of the second component in (\ref{p2}):
\begin{enumerate}
\item Regime I: The existence of a positive steady mutant population in (\ref{p2}) demands the following condition between the ecological parameters
\begin{equation}
\frac{\beta}{\sigma} > 1+ \frac{1}{\alpha\epsilon_1} \hspace{0.5cm} \mbox{or equivalently} \hspace{0.5cm} \epsilon_1 > \frac{\sigma}{\alpha(\beta-\sigma)} \hspace{0.2cm}. \label{inequation02}
\end{equation}
Now, the constant solution (\ref{p2}) describes a scenario where the non-mutant prey species is absent in favor of a mutant prey population, which shares the habitat with the predator species. By comparing the components in (\ref{p1}) and (\ref{p2}) we conclude that the steady mutant prey population associated with the stationary point $P_2$ is always greater than the non-mutant prey population for the stationary point $P_1$ in this first regime. Analogously the predator population is less for the steady state $P_2$ than for the constant solution $P_1$.

\item Regime II: For the parameter range
\begin{equation}
1+ \frac{1}{\epsilon_1}<\frac{\beta}{\sigma} < 1+ \frac{1}{\alpha\epsilon_1} \hspace{0.3cm} \mbox{equivalently} \hspace{0.3cm} \frac{\sigma}{\beta-\sigma}<\epsilon_1 < \frac{\sigma}{\alpha(\beta-\sigma)} \hspace{0.2cm}, \label{inequation02}
\end{equation}
the stationary point $P_2$ (a constant solution of (\ref{system2})) loses its ecological meaning because it moves to the region $y<0$, where the snake population is negative.
\end{enumerate}
In sum, the stationary point set ${\cal S}$ of (\ref{system2}) is given by ${\cal S}_I=\{P_0,P_1,P_2,R_{12}\}$ for the Regime I and by ${\cal S}_{II}=\{P_0,P_1,R_{12}\}$ for the Regime II, where $R_{12}=\{R_{12}^{(\mu)} : 0\leq \mu \leq 1\}$. In Figure 1 the arrangement of the stationary points in the region $\mathbb{E}$ is illustrated for the two previously introduced regimes. Notice that the point $P_2$ does not arise in the Regime II.

\begin{figure}[h]
\centerline{\includegraphics[height=4cm]{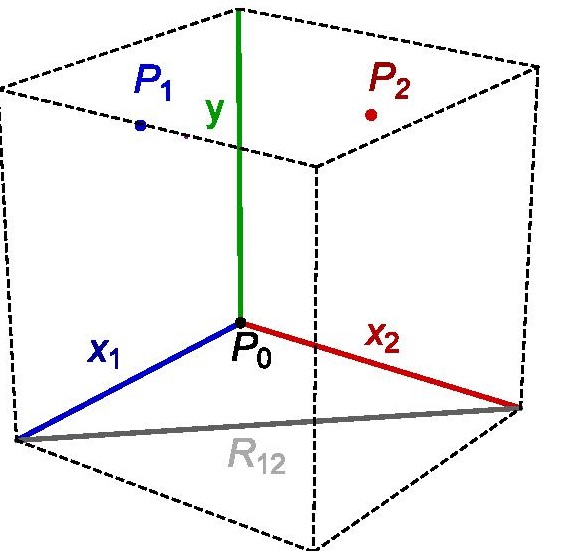} \hspace{2cm}
\includegraphics[height=4cm]{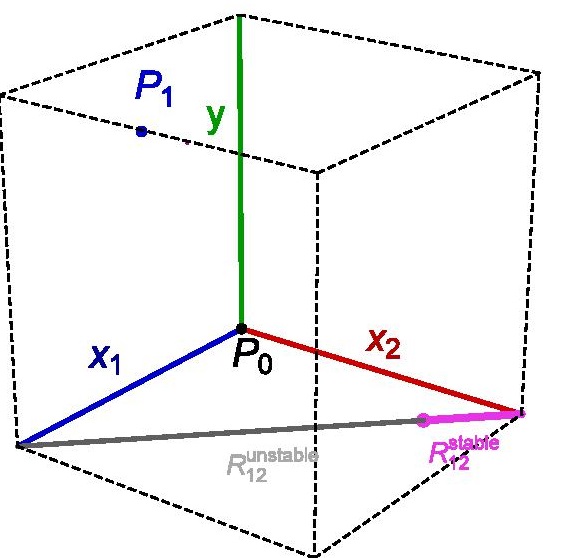}}
\caption{Distribution of stationary points in the octant $\mathbb{E}$ for the regime I (left) and for the regime II (right). For this last regime the line of degenerate stationary points $R_{12}$ is partitioned in two parts on the basis of stability criteria. }
\end{figure}

\vspace{0.2cm}

\noindent \textsc{Proposition 1:} \textit{The local stability of the stationary point set ${\cal S}\subset \mathbb{E}$ of the system (\ref{system2}) is described as follows:}

\vspace{0.2cm}

\noindent (A) \textit{For the regime I:}

\begin{enumerate}
\item \textit{The point $P_0$ and the set $R_{12}$ are unstable.}
\item \textit{$P_1$ is unstable with respect to $x_2$-fluctuations. If $\epsilon_1< \frac{\sigma+\beta}{\beta-\sigma}$ then $P_1$ is stable with respect to $x_1-y$ fluctuations but if $\epsilon_1>\frac{\sigma+\beta}{\beta-\sigma}$ then $P_1$ becomes unstable and an unique limit cycle arises lying in the $x_1-y$ plane.}
\item \textit{$P_2$ is stable with respect to $x_1$-fluctuations. If $\epsilon_1< \frac{\sigma+\beta}{\alpha(\beta-\sigma)}$ then $P_2$ is stable with respect to $x_2-y$ fluctuations but if $\epsilon_1>\frac{\sigma+\beta}{\alpha(\beta-\sigma)}$ then $P_2$ becomes unstable and an unique limit cycle arises lying in the $x_2-y$ plane.}
\end{enumerate}

\vspace{0.2cm}

\noindent (B) \textit{For the regime II:}
\begin{enumerate}
\item \textit{$P_0$ is unstable.}
\item \textit{The points $R_{12}^{(\mu)}$ are stable with respect to the $y$-fluctuations if $\mu>\mu_0$ and unstable if $\mu<\mu_0$ where}
    \begin{equation}
    \mu_0=\frac{\alpha}{1-\alpha} \Big[\frac{\sigma}{\alpha\epsilon_1(\beta-\sigma)}-1\Big] \hspace{0.3cm}. \label{mu0}
    \end{equation}
\item \textit{$P_1$ is unstable with respect to $x_2$-fluctuations. $P_1$ is stable with respect to $x_1-y$ fluctuations if $\epsilon_1< \frac{\sigma+\beta}{\beta-\sigma}$ but if $\epsilon_1>\frac{\sigma+\beta}{\beta-\sigma}$ then $P_1$ becomes unstable and an unique limit cycle arises lying in the $x_1-y$ plane.}
\end{enumerate}
\textsc{Proof:} See Appendix A.

\vspace{0.2cm}

Proposition 1 has proved that the non-mutant prey and predator population described by the steady state $P_1$ is locally unstable in our model when a mutant population is introduced. It will be checked that this is not only a local behavior restricted to the stationary point $P_1$ but it is a global pattern for any non-mutant prey and predator population. In order to justify this assertion the model dynamics will be expressed in the system of cylindrical coordinates
\begin{equation}
x_1 = \rho \cos \theta \hspace{0.3cm} ,\hspace{0.3cm} \ x_2 = \rho \sin \theta \hspace{0.3cm} , \hspace{0.3cm} y\equiv  y \hspace{0.3cm}. \label{cilindricas}
\end{equation}
The radial coordinate $\rho$ is proportional to the root mean square of the global prey population while the angular coordinate $\theta$ measures the ratio between the mutant and non-mutant populations. Now, the region with ecological relevance is defined by the cylindrical coordinate range
\[
\mathbb{E}=\{(\rho,\theta,y)\in \mathbb{R}^3 : \rho\geq 0, \,\, \theta\in [0,\pi/2],\,\, z\geq 0\} \hspace{0.3cm}.
\]
The use of (\ref{cilindricas}) turns (\ref{system2}) into the following differential equations
\begin{eqnarray}
\frac{d\rho}{d\tau} & =&  \rho \, \left[ 1-\rho ( \cos \theta+ \sin \theta )\right] - \frac{y \, \rho \, [1-(1-\alpha) \sin^2\theta]}{1+ \rho\,\epsilon_1 (\cos \theta + \alpha \sin \theta)} \hspace{0.3cm}, \nonumber \\
\frac{d \theta}{d\tau} & =&  \frac{(1-\alpha) \, \sin \theta\, \cos \theta\, \rho \, y}{1+ \rho \,\epsilon_1  (\cos \theta + \alpha  \sin \theta)} \hspace{4.8cm}, \label{cilisystem} \\
\frac{dy}{d\tau}&=& (\beta-\sigma)\, y\, -\, \frac{\beta\, y}{1+\rho\, \epsilon_1(  \cos \theta + \alpha  \sin \theta)} \hspace{2.5cm}. \nonumber
\end{eqnarray}
The second relation in (\ref{cilisystem}) gives the temporal derivative of the coordinate $\theta$ in terms of the ecological variables. It can be checked that this magnitude is always positive in the octant $\mathbb{E}$:
\[
\frac{d\theta}{d\tau} >0 \hspace{0.9cm} \forall \theta \in \left(0,\frac{\pi}{2}\right), \hspace{0.3cm} \alpha\in (0,1) \hspace{0.5cm}.
\]
This means that the dynamics favors the mutant prey population settlement. In our context the dextral snails are gradually replaced by the sinistral variant. Indeed, by inspecting this equation in more detail the factors which improve this behaviour can be identified. For example, if the asymmetry index is large (which involve a small value for $\alpha$) then the non-mutant to mutant population transition is speeded up. This process is intensified for large populations and when the number of non-mutant and mutant preys is similar although slows down when the population distribution is near to a coordinate plane.

\vspace{0.2cm}

Now, the nature of the limit sets associated to the system (\ref{system2}) will be employed to establish a classification of distinct scenarios involved in our model. This scheme splits up the parameter space into two regions delimitated by the condition $\alpha=\frac{\sigma}{\sigma+\beta}$. The identity of the limit sets for each of these two regions $\alpha>\frac{\sigma}{\sigma+\beta}$ and $\alpha<\frac{\sigma}{\sigma+\beta}$ is respectively established in Table 2 and 3. Notice that the ecological condition (\ref{range1}) involves that the results displayed in Table 3 only arise for cases where the snake specialization is strong, $\alpha< \frac{1}{2}$.

\begin{table}[h]
{\scriptsize
\begin{tabular}{|c|c|c|c|c|} \hline
$\alpha> \frac{\sigma}{\sigma+\beta}$ & Regime II & \multicolumn{3}{|c|}{\rule[-0.4cm]{0cm}{1cm} Regime I} \\ \hline\hline
$\epsilon_1$ \rule[-0.4cm]{0cm}{1cm} & $\Big( \frac{\sigma}{\beta-\sigma}, \frac{\sigma}{\alpha(\beta-\sigma)}\Big)$ & $\Big(\frac{\sigma}{\alpha(\beta-\sigma)}, \frac{\sigma+\beta}{\beta-\sigma}\Big)$  & $\Big( \frac{\sigma+\beta}{\beta-\sigma}, \frac{\sigma+\beta}{\alpha(\beta-\sigma)}\Big)$ & $\Big(\frac{\sigma+\beta}{\alpha(\beta-\sigma)},\infty\Big)$ \\ \hline\hline
\rule[-0.4cm]{0cm}{1cm} $\alpha$ limit set & Stationary point $P_1$ & Stationary point $P_1$ & $x_1-y$ limit cycle & $x_1-y$ limit cycle \\ \hline
\rule[-0.4cm]{0cm}{1cm} $\omega$ limit set & Point in $R_{12}^{\rm stable}$ & Stationary point $P_2$ & Stationary point $P_2$ & $x_2-y$ limit cycle \\ \hline
\rule[-0.4cm]{0cm}{1cm} & see Figure 3(a) & see Figure 2(a) & see Figure 2(b) & see Figure 2(c) \\ \hline
\end{tabular}}
\caption{Limit sets for the different parameter values $\epsilon_1$ in the case $\alpha> \frac{\sigma}{\sigma+\beta}$.}
\end{table}

\begin{table}[h]
{\scriptsize
\begin{tabular}{|c|c|c|c|c|} \hline
$\alpha< \frac{\sigma}{\sigma+\beta}$ & \multicolumn{2}{|c|}{\rule[-0.4cm]{0cm}{1cm} Regime II } & \multicolumn{2}{|c|}{\rule[-0.4cm]{0cm}{1cm} Regime I } \\ \hline\hline
$\epsilon_1$ \rule[-0.4cm]{0cm}{1cm}  & $\Big(\frac{\sigma}{\beta-\sigma}, \frac{\sigma+\beta}{\beta-\sigma} \Big)$ & $\Big( \frac{\sigma+\beta}{\beta-\sigma},\frac{\sigma}{\alpha(\beta-\sigma)} \Big)$ & $\Big(\frac{\sigma}{\alpha(\beta-\sigma)} ,\frac{\sigma+\beta}{\alpha(\beta-\sigma)} \Big)$ & $\Big(\frac{\sigma+\beta}{\alpha(\beta-\sigma)} ,\infty\Big)$ \\ \hline\hline
\rule[-0.4cm]{0cm}{1cm} $\alpha$ limit set & Stationary point $P_1$ & $x_1-y$ limit cycle &  $x_1-y$ limit cycle & $x_1-y$ limit cycle \\ \hline
\rule[-0.4cm]{0cm}{1cm} $\omega$ limit set & Point in $R_{12}^{\rm stable}$ & Point in $R_{12}^{\rm stable}$ & Stationary point $P_2$ & $x_2-y$ limit cycle \\ \hline
\end{tabular}}
\caption{Limit sets for different parameter values $\epsilon_1$ in the case $\alpha< \frac{\sigma}{\sigma+\beta}$.}
\end{table}

\noindent The local stability study, stated in Proposition 1, together with the global rule $\frac{d\theta}{d\tau}>0$ for $\theta\in (0,\frac{\pi}{2})$ allow us to accomplish a complete qualitative analysis of the solutions for our mathematical model. This description also illustrates the information included in Table 2 y 3. At this point, it is worth to notice that the model dynamics in absence of the predator population is restricted to the $x_1-x_2$ plane. For this particular case, characterized by the differential equations (\ref{particular2}), the solutions describe straight line orbits which asymptotically collapse in the line $R_{12}$. In Regime I, all the stationary points which comprise this set are unstable. A qualitative description of the solutions in the Regime I complying with the initial conditions assumed in this paper (where the mutant prey population is small) is provided in the following paragraphs.
\begin{enumerate}
\item For $\alpha>\frac{\sigma}{\sigma+\beta}$, the orbits in the phase space can be classified as follows:
\begin{enumerate}
\item If $\epsilon_1\in (\frac{\sigma}{\alpha(\beta-\sigma)}, \frac{\sigma+\beta}{\beta-\sigma})$ then the orbits initially approach to the stationary point $P_1$ for very small mutant prey populations, move away from the $x_1-y$ plane and finally tends to the stationary point $P_2$ in the $x_2-y$ plane, which is stable for this parameter range (see Figure 2(a)).

\item If $\epsilon_1 \in ( \frac{\sigma+\beta}{\beta-\sigma}, \frac{\sigma+\beta}{\alpha(\beta-\sigma)})$ then the orbits begin near the $x_1-y$ plane asymptotically describing a limit cycle confined to this plane although the increasing non-mutant prey population obliges the orbit to gradually approach to the stationary point $P_2$ (see Figure 2(b)).

\item If $\epsilon_1 \in (\frac{\sigma+\beta}{\alpha(\beta-\sigma)},\infty )$ then orbits evolve asymptotically from a limit cycle in the $x_1-y$ plane to a limit cycle in the $x_2-y$ plane. In our ecological context the snail and snake populations never stop oscillating while the sinistral variant is replacing the dextral one (see Figure 2(c)).
\end{enumerate}
\item For $\alpha<\frac{\sigma}{\sigma+\beta}$, the solution orbits exhibit a similar behaviour than those described in the previous point although now the case 1(a) does not arise. In other words, the initial non-mutant prey and predator populations are initially attracted by the limit cycle living in the $x_1-y$ plane but the final tendency of the orbits depends on the parameter $\epsilon_1$:
\begin{enumerate}
    \item If $\epsilon_1 \in ( \frac{\sigma}{\alpha(\beta-\sigma)}, \frac{\sigma+\beta}{\alpha(\beta-\sigma)})$ then the orbits asymptotically approach to the stationary point $P_2$ as the mutant prey population is settling the ecosystem (see Figure 2(b)).

    \item If $\epsilon_1 \in (\frac{\sigma+\beta}{\alpha(\beta-\sigma)},\infty )$ then orbits asymptotically evolve to the limit cycle placed in the $x_2-y$ plane, where the mutant prey and predator populations oscillate (see Figure 2(c)).
    \end{enumerate}
\end{enumerate}

\noindent It can be observed from our previous results that the situation in which there exist orbits asymptotically coming from the stationary point $P_1$ and evolving to a limit cycle in the $x_2-y$ plane is forbidden.

\begin{figure}[h]
\centerline{\includegraphics[height=4cm]{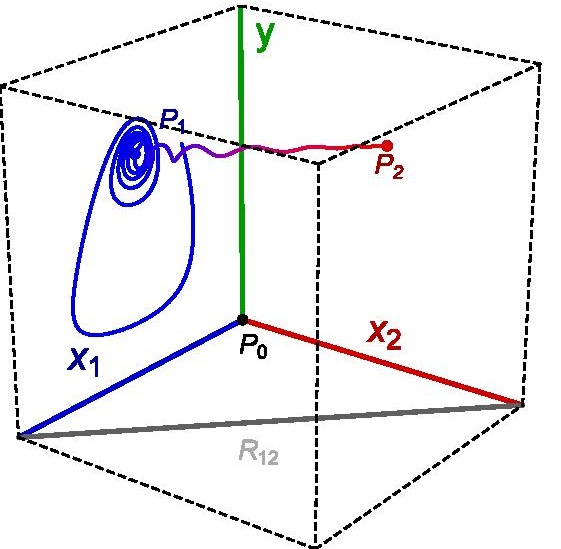} \hspace{0.5cm} \includegraphics[height=4cm]{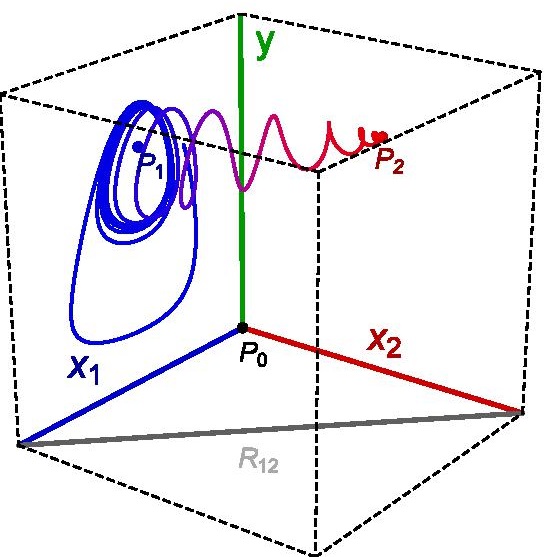} \hspace{0.5cm} \includegraphics[height=4cm]{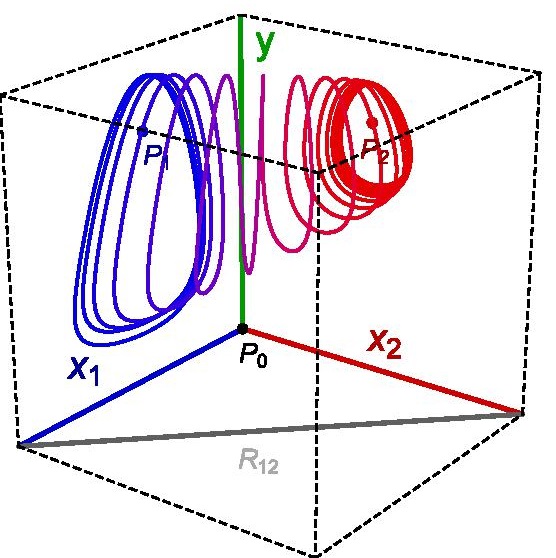}}
\caption{Orbits in the octant $\mathbb{E}$ for Regime I with $\alpha>\frac{\sigma}{\sigma+\beta}$  for: (a) $\epsilon_1\in (\frac{\sigma}{\alpha(\beta-\sigma)}, \frac{\sigma+\beta}{\beta-\sigma})$, (b) $\epsilon_1 \in ( \frac{\sigma+\beta}{\beta-\sigma}, \frac{\sigma+\beta}{\alpha(\beta-\sigma)})$ and (c) $\epsilon_1 \in (\frac{\sigma+\beta}{\alpha(\beta-\sigma)},\infty )$.}
\end{figure}

\vspace{0.2cm}

In Regime II a new pattern in the behavior of the solution orbits is found. In this case, the stationary point $P_2$, which describes the coexistence of steady mutant prey and predator populations, leaves the ecological region $\mathbb{E}$. Despite of this fact, there exists a dense orbit set confined to the $x_2-y$ plane by the dynamics ruled by the equations (\ref{particular1b}), which tends to the steady state $(0,1,0)$. This point is the intersection between the line $R_{12}$ and the previously mentioned plane. Therefore, any other orbit described in the interior of the octant $\mathbb{E}$ cannot cross this plane. All these facts allow us to conclude that the orbits beginning near the $x_1-y$ plane must asymptotically approach to a stable stationary point in the segment
\[
R_{12}^{\rm stable}= \{(\mu,1-\mu,0) : \mu\in (0,\mu_0)\}
\]
where $\mu_0$ is given by (\ref{mu0}). In this case only the non-mutant and mutant prey populations coexist. In this scenario the predator species becomes extinct. The exact limit point of this type of orbits depends on the initial conditions. Taking into account the previous results, the classification of the orbits in the Regime II for our mathematical model can be described as follows:
\begin{enumerate}
\item For $\alpha>\frac{\sigma}{\sigma+\beta}$ then $\epsilon_1\in (\frac{\sigma}{\beta-\sigma}, \frac{\sigma}{\alpha(\beta-\sigma)})$ must be necessarily verified in the regime II. Here the orbits tend to approach to the stationary point $P_1$ when the mutant prey population is initially introduced. The orbits move away from the $x_1-y$ plane as the predator population decreases until extinction. In our ecological context the orbits asymptotically tend to a coexistence population between sinistral and dextral snails. This behavior resembles the orbit displayed in Figure 3(a).

\item For $\alpha<\frac{\sigma}{\sigma+\beta}$ two different situations must be distinguish:
\begin{enumerate}
    \item If $\epsilon_1 \in ( \frac{\sigma}{\beta-\sigma}, \frac{\sigma+\beta}{\beta-\sigma})$ then the orbits behave in a similar way than those described in the previous point (see Figure 3(a)).

    \item If $\epsilon_1 \in (\frac{\sigma+\beta}{\beta-\sigma},\frac{\sigma}{\alpha(\beta-\sigma)})$ then orbits tend to initially follow a limit cycle when the mutant species is minority and as before the predator species asymptotically extinguishes giving rise to coexistence between the mutant and non-mutant prey populations (see Figure 3(b)).
    \end{enumerate}
\end{enumerate}

\begin{figure}[h]
\centerline{\includegraphics[height=4cm]{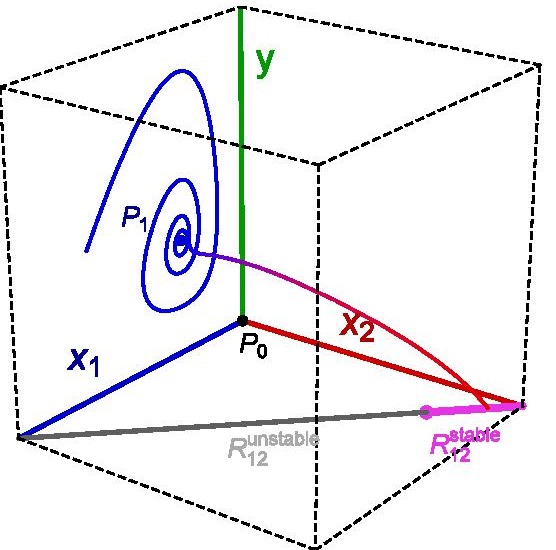} \hspace{1.5cm} \includegraphics[height=4cm]{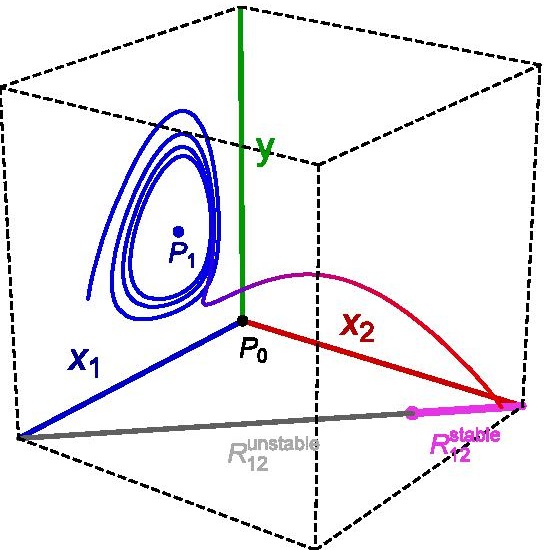}}
\caption{Orbits in the phase plane in Regime II with $\alpha<\frac{\sigma}{\sigma+\beta}$ for (a) $\epsilon_1\in (\frac{\sigma}{\beta-\sigma}, \frac{\sigma+\beta}{\beta-\sigma})$ and (b) $\epsilon_1 \in ( \frac{\sigma+\beta}{\beta-\sigma}, \frac{\sigma}{\alpha(\beta-\sigma)})$.}
\end{figure}

\section{Conclusions}

The interactions between a dextral snail-eating specialist \textit{Pareas} snake population, a settled dextral snail population and a (mutant) sinistral snail population have been analyzed by constructing a mathematical model, which is described by the system of ordinary differential equations (\ref{system}). This model assumes a logistic growth for the total snail population. Obviously, the two snail variants are characterized by the same ecological parameters. The construction of the expressions (\ref{deltas}) and (\ref{func}), which describe the hunting habits followed by \textit{Pareas} snakes, supposes than the predator spends a constant time period a day searching, hunting and handling preys (following the same hypotheses employed in the deduction of Holling type II functional responses, see \cite{Holling1, Holling2, Holling3, May, Maynard}). The mandibular asymmetry presented by \textit{Pareas} snakes (an evolutionary adaptation to the feeding on dextral snails) implies that hunting efficiencies on dextral/sinistral snails are different and offers survival advantage to the sinistral snail variant. This, in turn, introduces an asymmetry in the equations (\ref{system}), which is measured by the parameter $\alpha$. This parameter can be assessed for different \textit{Pareas} snake species ranging from the extreme value $\alpha=0.12$ for \textit{Pareas Iwasakii} to $\alpha=1$ for snakes with symmetrical dentation. \textit{Pareas Carinatus} is an astonishing case where a weak dental asymmetry is offset by its skill to recognize snail handedness before striking the prey \cite{Danaisawadi2016}. As a consequence, the parameter $\alpha$ is approximately 1 for this species.

The distinct scenarios of evolution of the three populations are described in this paper. Evolutionary ecology researches indicates the conviction that in the previously described context the sinistral snail population will replace the dextral variant. This mechanism can be used to explain left-right reversal in snails that have happened several times on Earth \cite{Hoso1,Hoso2,Hoso3}. The analysis of the solutions for our model mostly supports this proposal but it also opens new possibilities for extreme cases such as the \textit{Pareas Iwasakii} snakes. From the qualitative analysis of (\ref{system}) the extinction of \textit{Pareas} snakes is also a possible picture when snake species exhibit high specialization. Here, \textit{Pareas Carinatus} teaches us the need to adapt to changes.

The model described in this paper can also be applied to other ecosystems where two prey variants coexist with a predator that exhibits a preference for feeding on one of the prey variants. In this paper, the snail chirality is the characteristic that determines this bias but other features, such as skin color, camouflage skills, etc., can play similar roles. The evolution study of invasive species populations that show similar characteristics than native species also constitutes a particularly interesting application of this model.

\section{ACKNOWLEDGEMENTS}

The authors acknowledge the Spanish Ministerio de Econom\'{\i}a y Competitividad for financial support under grant MTM2014-57129-C2-1-P. They are also grateful to the Junta de Castilla y Le\'on for financial help under grant VA057U16.





\appendix

\section{}

In this Appendix the proof of Proposition 1 is included. It involves standard and well established techniques, see \cite{Sugie, Gasull1, Gasull2, Seo, Sigmund, Kuang1990, Farkas1990} and references therein, which are applied to our context. The local stability of the stationary points for the differential equations (\ref{system2}) in the ecological region $\mathbb{E}$ is analyzed in the following points:

\vspace{0.2cm}

\noindent (1) For the trivial solution $P_0\equiv (0,0,0)$ the linear stability matrix is written as follows
\[
L[P_0]= \left( \begin{array}{ccc} 1 & 0 & 0\\ 0 & 1 & 0 \\ 0 &0 & -\sigma \end{array} \right) \hspace{0.3cm},
\]
which implies that $P_0$ corresponds to a saddle point. Therefore, $P_0$ is an unstable stationary point because $\sigma>0$.

\vspace{0.2cm}

\noindent (2) The linear stability of the point $P_1$ (describing steady coexistence between non-mutant prey and predator species) is governed by the eigenvalues of the matrix
\[
L[P_1]= \left( \begin{array}{ccc}
-\frac{\sigma(\beta+\sigma +\epsilon_1(\sigma-\beta))}{\beta \epsilon_1 (\beta -\sigma)} & \frac{\sigma}{\beta} [\alpha - \frac{\beta+\alpha \sigma}{(\beta-\sigma)\epsilon_1}] & -\frac{\sigma}{\beta \epsilon_1} \\ 0 & (1-\alpha)[1-\frac{\sigma}{(\beta-\sigma)\epsilon_1}] & 0 \\ (\beta-\sigma)\epsilon_1-\sigma & \alpha[(\beta-\sigma)\epsilon_1-\sigma] & 0
\end{array} \right) \hspace{0.3cm}.
\]
The characteristic polynomial of $L[P_1]$ reads
\[
\textstyle p_c(\lambda)= \Big( \lambda^2 + \frac{\sigma[\beta+\sigma+\epsilon_1(\sigma-\beta)]}{\beta \epsilon_1 (\beta-\sigma)} \lambda - \frac{\sigma(\sigma+\epsilon_1 \sigma-\beta \epsilon_1)}{\beta \epsilon_1}\Big)\Big( \lambda - (1-\alpha) [1-\frac{\sigma}{(\beta-\sigma)\epsilon_1}] \Big) \hspace{0.3cm}.
\]
The first factor of this polynomial rules the evolution of the fluctuations induced in the non-mutant and predator populations. As previously mentioned a Holling type II prey-predator model emerges when the dynamics is confined to this $x_1-y$ coordinate plane. We can now use the known results in this framework \cite{Sugie, Gasull1, Gasull2, Seo, Sigmund, Kuang1990, Farkas1990}. The value $\epsilon_1 = \frac{\sigma+\beta}{\sigma-\beta}$ characterizes the occurrence of a Hopf bifurcation. For $\epsilon_1 < \frac{\sigma+\beta}{\sigma-\beta}$ the point $P_1$ is stable with respect to fluctuations in the $x_1-y$ plane and becomes unstable if the condition $\epsilon_1 > \frac{\sigma+\beta}{\sigma-\beta}$ is verified. In this last case, the presence of an unique limit cycle is guaranteed. From an ecological point of view the previous results are summarized as follows:
\begin{enumerate}
\item[a.] If $\frac{\sigma}{\beta-\sigma}<\epsilon_1< \frac{\sigma+\beta}{\beta-\sigma}$ then the stationary point $P_1$ is stable, and the predator and prey populations tend to this steady solution if the mutant species is absent.

\item[b.] If $\epsilon_1> \frac{\sigma+\beta}{\beta-\sigma}$ and there is no mutant individuals then the non-mutant prey-predator population asymptotics is dominated by a cycle limit embedded in the $x_1-y$ plane.
\end{enumerate}

\noindent Finally, the eigenvalue
\[
\lambda_3=\frac{(\alpha-1)(\sigma+\epsilon_1(\sigma-\beta))}{\epsilon_1 (\beta-\sigma)} \hspace{0.3cm},
\]
extracted from the last factor of $p_c(\lambda)$, establishes the stability of the steady point $P_1$ when mutant prey variations are applied. Under our ecological assumptions (\ref{alpharange}), (\ref{range1}) and (\ref{inequation01}), the value of $\lambda_3$ is always positive. Therefore, the steady state $P_1$ is unstable when these type of perturbations are introduced.

\vspace{0.2cm}

\noindent (3) The stability of the steady point $P_2$ (which has ecological sense only in the Regime I) is analyzed in this point. The linear stability matrix reads
\[
L[P_2]= \left( \begin{array}{ccc} \frac{(1-\alpha)[\sigma+\alpha(\sigma-\beta)\epsilon_1]}{\alpha^2 \epsilon_1(\beta-\sigma)} & 0 & 0 \\ -\frac{\sigma[\alpha\beta+\sigma+\alpha(\sigma-\beta)\epsilon_1]}{\alpha^2 \beta \epsilon_1(\beta-\sigma)} & - \frac{\sigma[\beta+\sigma+\alpha\epsilon_1(\sigma-\beta)]}{\alpha \beta\epsilon_1(\beta-\sigma)} & -\frac{\sigma}{\beta \epsilon_1} \\ \frac{-\sigma+\alpha(\beta-\sigma)\epsilon_1}{\alpha^2} & -\frac{\sigma}{\alpha} +(\beta-\sigma)\epsilon_1 & 0
\end{array} \right) \hspace{0.3cm},
\]
which provides us with the characteristic polynomial
\[
\textstyle p_c(\lambda)= \Big( \lambda^2 + \frac{\sigma[\beta+\sigma +\alpha \epsilon_1(\sigma-\beta)]}{\alpha \beta \epsilon_1(\beta-\sigma)} \lambda - \frac{\sigma(\sigma+\alpha\epsilon_1 \sigma- \alpha \beta \epsilon_1)}{\alpha \beta \epsilon_1}\Big)\Big( \lambda - \frac{(1-\alpha)[\sigma+\alpha(\sigma-\beta)\epsilon_1]}{\alpha^2 \epsilon_1(\beta-\sigma)} \Big) \hspace{0.3cm}.
\]
The same arguments used in the stability study of the point $P_1$ can be applied in this case. It can be concluded that if the condition
\[
\epsilon_1 < \frac{\sigma+\beta}{\alpha(\sigma-\beta)}
\]
holds, the point $P_2$ is stable with respect to mutant prey and predator population fluctuations. On the other hand, a unique limit cycle lies in the $x_2-y$ plane for the parameter range determined by the inequality
\[
\epsilon_1 > \frac{\sigma+\beta}{\alpha(\sigma-\beta)} \hspace {0.3cm}.
\]
For the sake of ecological interpretation we distinguish the following two cases:
\begin{enumerate}
\item[a.] If $\frac{\sigma}{\alpha(\beta-\sigma)}< \epsilon_1< \frac{\sigma+\beta}{\alpha(\beta-\sigma)}$ then the stationary point $P_2$ is stable when fluctuations embedded in the $x_2-y$ plane are introduced. In the absence of the non-mutant population, the predator and mutant prey populations tend to this steady solution.

\item[b.] For the parameter range $\epsilon_1> \frac{\sigma+\beta}{\alpha(\beta-\sigma)}$, the predator-mutant prey populations shall evolve following a cycle limit confined to the $x_2-y$ plane.
\end{enumerate}

\noindent Finally, the stability of the stationary point $P_2$ against non-mutant prey population fluctuations is discussed. The third eigenvalue of the matrix $L(P_2)$,
\[
\lambda_3= \frac{(\alpha-1)(\sigma+\epsilon_2(\sigma-\beta))}{\alpha \epsilon_2 (\beta-\sigma)} \hspace{0.3cm},
\]
which determines the temporal evolution of this type of perturbations, is negative for the parameter values allowed in the regime I. This means that the point $P_2$ is stable when non-mutant population variations are applied.

\vspace{0.2cm}

\noindent (4) Now, we shall deal with the local stability study for the set of stationary points $R_{12}^{(\mu)} \equiv (\mu,1-\mu,0)$,  $\mu\in [0,1]$. These steady states describe the coexistence between the prey variants without the presence of the predator species. The linear stability matrix is given by
\[
L[R_{12}^{(\mu)}]= \left( \begin{array}{ccc} -\mu & -\mu & -\frac{\mu}{1+\epsilon_1 \mu + \epsilon_2(1-\mu)} \\ -(1-\mu) & -(1-\mu) & -\frac{\alpha(1-\mu)}{1+\epsilon_1 \mu + \epsilon_2(1-\mu)} \\ 0 & 0 & \beta-\sigma - \frac{\beta}{1+\epsilon_1 \mu + \epsilon_2(1-\mu)} \end{array} \right) \hspace{0.3cm},
\]
whose eigenvalues are expressed as
\[
\lambda_1=0\hspace{0.5cm},\hspace{0.5cm} \lambda_2= -1 \hspace{0.5cm} ,\hspace{0.5cm} \lambda_3= \beta - \sigma - \frac{\beta}{1+\epsilon_1 \mu + \epsilon_2(1-\mu)} \hspace{0.3cm} .
\]
The presence of a vanishing eigenvalue indicates the influence of non-linear terms in the local stability analysis. The following change of coordinates
\begin{eqnarray}
&& x_1-\mu= -\overline{x}_1+ \frac{\mu}{1-\mu} \overline{x}_2 +\Big[ -\frac{(1-\alpha)(1-\mu)\mu}{-\sigma+ \epsilon_1 (\beta-\sigma)[\mu+\alpha(1-\mu)]} + \nonumber \\ && \hspace{4cm} + \frac{\mu (\alpha \mu-\alpha-\mu)}{1-\sigma-\epsilon_1[\alpha(-1+\mu)-\mu](1+\beta-\sigma)} \Big] \overline{y} \hspace{0.3cm}, \nonumber \\
& & x_2-(1-\mu)=\overline{x}_1+ \overline{x}_2 +\Big[ \frac{(1-\alpha)(1-\mu)\mu}{-\sigma+ \epsilon_1 (\beta-\sigma)[\mu+\alpha(1-\mu)]} - \label{change} \\ && \hspace{4cm} - \frac{(-1+\mu) [\alpha(-1+\mu)-\mu]}{1-\sigma-\epsilon_1[\alpha(-1+\mu)-\mu](1+\beta-\sigma)} \Big] \overline{y}  \hspace{0.3cm}, \nonumber\\
&& y= \overline{y} \hspace{0.3cm}, \nonumber
\end{eqnarray}
places the stationary point $R_{12}^{(\mu)}$ at the origin and orients the coordinate axes along the principal directions of the stability matrix $L[R_{12}^{(\mu)}]$. The use of these new variables turns (\ref{system2}) into the equations
\begin{eqnarray}
\frac{d\overline{x}_1}{dt} &=& - \frac{1}{1-\mu} \overline{x}_1 \,\overline{x}_2 + A_1 \overline{x}_1 \, \overline{y} + A_2  \overline{x}_2 \overline{y} +  o^3(\overline{x}_1,\overline{x}_2,\overline{y}) \hspace{0.3cm}, \nonumber \\
\frac{d\overline{x}_2}{dt} &=&  - \overline{x}_2  + o^2(\overline{x}_1,\overline{x}_2,\overline{y}) \hspace{0.3cm}, \label{system3} \\
\frac{d\overline{y}}{dt} &=& \Big( \beta-\sigma - \frac{\beta}{1+\epsilon_1(\alpha+ \mu - \alpha \mu)} \Big) \overline{y}+ o^2(\overline{x}_1,\overline{x}_2,\overline{y})  \hspace{0.3cm}. \nonumber
\end{eqnarray}
The expressions on the right hand side of the system (\ref{system3}) have been written up to dominant order in the new variables $\overline{x}_1$, $\overline{x}_2$ and $\overline{y}$. Moreover, the notations
\begin{eqnarray}
A_1 &=& \frac{-\alpha \beta +(1-\alpha)^2 (-1+\mu)\mu \sigma}{\beta(-\alpha-\mu +\alpha\mu)(-1+\epsilon_1(-\alpha-\mu+\alpha\mu))} + \nonumber \\ &+& \frac{(\beta-\sigma)(1-\alpha)^2(-1+\mu)\mu\sigma}{\beta(-\alpha-\mu+\alpha\mu)(-\sigma+\epsilon_1 (-\beta+\sigma)[\alpha(-1+\mu)-\mu])} + \label{a1a2} \\ &+&\frac{\alpha+\mu -\alpha\mu}{1-\sigma-\epsilon_1(1+\beta-\sigma)[\alpha(-1+\mu)-\mu]} \hspace{0.3cm}, \nonumber \\[0.2cm]
A_2&=& \frac{(-1+\alpha)\mu\sigma}{\beta[-1+\epsilon_1(\alpha(-1+\mu)-\mu)]} + \frac{(-1+\alpha)\mu(\beta+\beta\sigma-\sigma^2)}{\beta[-\sigma +(\alpha(-1+\mu)-\mu)\epsilon_1(-\beta+\sigma)]} \nonumber
\end{eqnarray}
have been introduced in (\ref{system3}). From (\ref{system3}) it can be concluded that the points $R_{12}^{(\mu)}$ are stable with respect to $\overline{x}_2$-fluctuations. At first order, this type of fluctuations follows the form
\[
\overline{x}_2 = \overline{x}_2(0) \cdot e^{-t} \hspace{0.3cm},
\]
where $\overline{x}_2(0)$ denotes the initial perturbation. The linear approximation introduced in the third equation of (\ref{system3}) rules the local stability of the points $R_{12}^{(\mu)}$ with respect to predator population fluctuations. At first order, the expression
\begin{equation}
\overline{y} = \overline{y}(0) \cdot e^{[\beta-\sigma - \frac{\beta}{1+\epsilon_1(\alpha+ \mu - \alpha \mu)}]\,t} =  \overline{y}(0) \, e^{\lambda_3 \,t} \label{lambda3enr12}
\end{equation}
determines the evolution of these fluctuations. As before, $\overline{y}(0)$ denotes the initial predator population variation. The sign of the eigenvalue $\lambda_3$ introduced in (\ref{lambda3enr12}) distinguishes two different pieces of the line $R_{12}=\{R_{12}^{(\mu)}:0\leq \mu\leq 1 \}$ where the stability behavior changes. The threshold value
\begin{equation}
\mu_0=\frac{\alpha}{1-\alpha} \Big[ \frac{\sigma}{\alpha\epsilon_1(\beta-\sigma)} -1 \Big] \label{mu0}
\end{equation}
of the parameter $\mu$ delimitates the previously mentioned segments. In more detail, the set $R_{12}^{\rm stable}= \{R_{12}^{(\mu)}: \mu\in (0,\mu_0) \}$ (located on $R_{12}$) comprises stable stationary points whereas the set $R_{12}^{\rm unstable}= \{R_{12}^{(\mu)}: \mu\in (\mu_0,1) \}$ corresponds to unstable stationary points against predator population fluctuations. From an ecological point of view, the previous situation (with presence of stable and unstable stationary points in the set $R_{12}$) occurs only in the regime II (described in Section 4). The bisection of $R_{12}$ at a point in the region $\mathbb{E}$ demands that $\mu_0\in (0,1)$. The condition $\mu_0 >0$ leads to the constraint $\frac{\sigma}{\alpha(\beta-\sigma)}>\epsilon_1$ whereas $\mu_0<1$ involves that $\frac{\sigma}{\beta-\sigma}<\epsilon_1$. This parameter range defines the regime II, see formula (\ref{inequation02}). On the other hand, all the steady states belonging to $R_{12}$ are unstable with respect to predator population fluctuations in Regime I.

\vspace{0.2cm}

Now, the first differential equation in (\ref{system3})
\[
\frac{d\overline{x}_1}{dt} = \Big[ A_1  \,  \overline{y}(0)\, e^{\lambda_3 \,t}- \frac{\overline{x}_2(0)\, e^{-t}}{1-\mu}\Big] \overline{x}_1 + A_2 \, \overline{x}_2(0)\,  \overline{y}(0)\, e^{(\lambda_3-1) \,t}
\]
rules the temporal evolution of the $\overline{x}_1$-fluctuations at the dominant order. If the initial predator population is zero the steady states $R_{12}$ are stable with respect to prey population perturbations.

\vspace{0.2cm}

For the sake of completeness, the extreme points of the line $R_{12}$ are analyzed separately:

\vspace{0.2cm}

\noindent (4a) For the stationary point $R_{12}^{(\mu=1)}\equiv (1,0,0)$, the change of coordinates (\ref{change}) reduces to the form
\[
x_1-1= -\overline{x}_1+ \overline{x}_2 +\frac{\overline{y}}{\epsilon_1(-\beta+\sigma-1)+\sigma-1} \hspace{0.3cm},\hspace{0.3cm}
x_2=\overline{x}_1 \hspace{0.3cm},\hspace{0.3cm}
y= \overline{y} \hspace{0.3cm},
\]
and (\ref{system3}) becomes
\begin{eqnarray*}
\frac{d\overline{x}_1}{dt} &=& - \overline{x}_1 \,\overline{x}_2 - \Big[ \frac{\alpha}{1+\epsilon_1} + \frac{1}{-1+\sigma+\epsilon_1(-1-\beta+\sigma)} \Big] \overline{x}_1 \overline{y} + o^3(\overline{x}_1,\overline{x}_2,\overline{y})  \hspace{0.3cm},\\
\frac{d\overline{x}_2}{dt} &=&  - \overline{x}_2  + o^2(\overline{x}_1,\overline{x}_2,\overline{y})  \hspace{0.3cm},\\
\frac{d\overline{y}}{dt} &=& \Big( -\sigma + \frac{\beta \epsilon_1}{1+\epsilon_1} \Big) \overline{y}+ o^2(\overline{x}_1,\overline{x}_2,\overline{y})  \hspace{0.3cm}.
\end{eqnarray*}
In this case, the stability condition with respect to predator fluctuations $-\sigma+\frac{\beta \epsilon_1}{1+\epsilon_1}<0$ implies that $\epsilon_1<\frac{\sigma}{\beta-\sigma}$. This requirement is never verified because of the ecological parameter restriction (\ref{inequation01}). Therefore, the stationary point $R_{12}^{(\mu=1)}$ is unstable even in the regime II.

\vspace{0.2cm}

\noindent (4b) For the particular stationary point $R_{12}^{(\mu=0)}\equiv (0,1,0)$, the changes of coordinates (\ref{change}) reads
\[
x_1= -\overline{x}_1 \hspace{0.3cm},\hspace{0.3cm} x_2-1=\overline{x}_1 + \overline{x}_2 + \frac{\alpha}{\alpha\epsilon_1(-\beta+\sigma-1)+\sigma-1} \overline{y}\hspace{0.3cm},\hspace{0.3cm}  y= \overline{y} \hspace{0.3cm},
\]
which turns (\ref{system3}) into
\begin{eqnarray*}
\frac{d\overline{x}_1}{dt} &=& - \overline{x}_1 \,\overline{x}_2 - \Big[ \frac{1}{1+\alpha \epsilon_1} + \frac{\alpha}{-1+\sigma+\alpha \epsilon_1(-1-\beta+\sigma)} \Big] \overline{x}_1 \overline{y} + o^3(\overline{x}_1,\overline{x}_2,\overline{y}) \\
\frac{d\overline{x}_2}{dt} &=&  - \overline{x}_2  + o^2(\overline{x}_1,\overline{x}_2,\overline{y})\hspace{0.3cm},\\
\frac{d\overline{y}}{dt} &=& \Big( \beta-\sigma - \frac{\beta}{1+\alpha \epsilon_1} \Big) \overline{y}+ o^2(\overline{x}_1,\overline{x}_2,\overline{y}) \hspace{0.3cm} .
\end{eqnarray*}
Stability condition against the predator fluctuations $\beta-\sigma-\frac{\beta}{1+\alpha\epsilon_1}<0$ involves that $\epsilon_1<\frac{\sigma}{\alpha(\beta-\sigma)}$. Therefore, the stationary point $R_{12}^{(\mu=0)}$ is always stable in Regime II.

\end{document}